\documentclass[%
 reprint,
 amsmath,amssymb,
 aps,
]{revtex4-1}
\usepackage[latin1]{inputenc}
\usepackage{mathtools}
\usepackage{slashed}
\usepackage{physics}
\usepackage{array}
\usepackage{amsmath}
\usepackage{xcolor}
\usepackage{amsfonts}
\usepackage{amssymb}
\usepackage{relsize}
\usepackage{graphicx}
\usepackage{amssymb,epsfig,subfigure}
\usepackage{hyperref}
\usepackage{comment}
\usepackage{tabularx}
\usepackage{bm}
\usepackage{euscript}
\usepackage{graphicx}
\usepackage{mathrsfs}
\usepackage[scr=boondox]{mathalfa}
\usepackage{color}
\usepackage [english]{babel}
\usepackage [autostyle, english = american]{csquotes}
\MakeOuterQuote{"}
\usepackage{amsfonts}
\usepackage{exscale}
\usepackage{amsbsy}
\usepackage{subfigure}
\usepackage{textcomp}
\usepackage{comment}
\usepackage{hyperref}
\usepackage{slashed}
\usepackage{tabularx}
\usepackage{euscript}
\usepackage{graphicx}
\usepackage{color}
\usepackage{amsfonts}
\usepackage{exscale}
\usepackage{amsbsy}
\usepackage{subfigure}
\usepackage{textcomp}
\usepackage{comment}
\usepackage{hyperref}
\usepackage{bm}
\usepackage{wrapfig}
\usepackage[font=footnotesize,labelfont=bf]{caption}

\newcommand{\beq}{\begin{eqnarray}}
\newcommand{\eeq}{\end{eqnarray}}
\begin{document}

\def\ppnumber{\vbox{\baselineskip14pt
}}

\def\ppdate{
} \date{\today}

\title{\bf Crystalline Responses for Rotation-Invariant Higher-Order Topological Insulators}
\author{Julian May-Mann}
\affiliation{ \it Department of Physics and Institute for Condensed Matter Theory,\\  \it University of Illinois at Urbana-Champaign, \\  \it 1110 West Green Street, Urbana, Illinois 61801-3080, USA}
\author{Taylor L. Hughes}
\affiliation{ \it Department of Physics and Institute for Condensed Matter Theory,\\  \it University of Illinois at Urbana-Champaign, \\  \it 1110 West Green Street, Urbana, Illinois 61801-3080, USA}

\begin{abstract}
Two-dimensional higher-order topological insulators can display a number of exotic phenomena such as half-integer charges localized at corners or disclination defects. In this paper, we analyze these phenomena, focusing on the paradigmatic example of the quadrupole insulator with $C_4$ rotation symmetry, and present a topological field theory description of the mixed geometry-charge responses. Our theory provides a unified description of the corner and disclination charges in terms of a physical geometry (which encodes disclinations), and an effective geometry (which encodes corners). We extend this analysis to interacting systems, and predict the response of fractional quadrupole insulators, which exhibit charge $e/2(2k+1)$ bound to corners and disclinations.
\end{abstract}

\maketitle

\bigskip
\newpage

Higher-order topological insulators (HOTIs) are forms of crystalline topological phases that host gapped surfaces which are connected by gapless corners or hinges. The first example of a HOTI was presented in Ref. \onlinecite{benalcazar2017quantized}, where it was shown that there exists a $C_4$ symmetric HOTI in $2$D (two spatial dimensions) that hosts half-integer corner charges when defined on a lattice with boundary. This insulator has vanishing electric polarization, but harbors a non-vanishing $xy$ quadrupole moment, and hence has been dubbed a topological quadrupole insulator (QI). In addition, it has been shown that this insulating phase is sensitive to the presence of $\pi/2$ disclinations on which half-integer charges are bound\cite{li2020fractional}. After this initial development the family of HOTIs has expanded to include other $2$D HOTIs with corner modes, and various $3$D HOTIs with either hinge modes or corner modes\cite{benalcazar2017electric,song2017d,langbehn2017reflection,schindler2018higher,khalaf2018higher,liu2019shift,benalcazar2019quantization,cualuguaru2019higher}. There have also been a number of experimental realizations of some classes of  HOTIs\cite{noh2018topological,serra2018observation,peterson2018quantized,imhof2018topolectrical,xue2019acoustic,zhang2019second,ni2019observation,noguchi2021evidence,schindler2018higherB,aggarwal2021evidence}.

Despite the rapid advances, our understanding of HOTIs still pales in comparison to our understanding of (first-order) topological insulators. While a number of detailed works analyze and classify HOTIs based on their symmetries\cite{song2017topological,song2017interaction,thorngren2018gauging, rasmussen2020classification}, the field theoretic understanding of HOTI responses is incomplete (see Ref. \onlinecite{you2021} for the topological response of subsystem symmetry protected HOTIs). Historically topological response theories have been a powerful tool with which to probe the physics of topological insulators, and, in turn, topological insulators have helped to provided new contexts in which one can realize topological field theories\cite{haldane1988,qi2008topological}. 
Motivated by this, we consider the response properties of the QI, especially its corner and disclination bound charge responses. Although we only consider the QI with $C_4$ symmetry here, our analysis can be straightforwardly generalized to other $C_n$ symmetric HOTIs that have a Dirac fermion description in the continuum, and perhaps even to other cases.

To analyze the $C_4$ symmetric QI we will consider a 4-band model of spinless fermions on a square lattice at half filling. The Bloch Hamiltonian is:
\begin{equation}\begin{split}
h^{q}(\vec{k}) = &\sin(k_x)\Gamma^1  + \sin(k_y)\Gamma^2 + \Delta_1 \Gamma^3\\
&+\tfrac{1}{\sqrt{2}}\left[\Delta_2 + \cos(k_x)-\cos(k_y)\right]\Gamma^4 \\&+\tfrac{1}{\sqrt{2}}\left[\Delta_3 + \cos(k_x)+\cos(k_y)\right]\Gamma^0,
\label{eq:QILatticeModel}\end{split}\end{equation}
where, $\Gamma^1 = \sigma^3 \otimes \sigma^2$, $\Gamma^2 = -\sigma^3 \otimes \sigma^1$, $\Gamma^3 = \sigma^1 \otimes \sigma^0$, $\Gamma^4 = \sigma^2 \otimes \sigma^0$, $\Gamma^0 = \sigma^3\otimes \sigma^3,$  $\sigma^{1,2,3}$ are Pauli matrices, and $\sigma^0$ is the $2\times 2$ identity. As shown in Appendix \ref{app:DiffQuad}, this Hamiltonian is equivalent, up to a unitary transformation, to the original model presented in Ref. \onlinecite{benalcazar2017quantized}. Before considering the continuum limit, we first recall some known features of this model (a more detailed analysis can be found in Ref. \onlinecite{benalcazar2017electric}). The energy bands derived from Eq. \ref{eq:QILatticeModel} are doubly degenerate, and there is a gap between the upper and lower pairs of bands which closes when $(\Delta_1,\Delta_2,\Delta_3) = (0,0,\pm 2)$, or $(0,\pm 2,0)$. The Bloch Hamiltonian has a (spinless) time-reversal symmetry $\mathcal{T} = \Gamma^2\Gamma^4 \mathcal{K}$ ($\mathcal{K}$ is complex conjugation), and when $\Delta_1 = \Delta_2 = 0$ there is a $C_4$ rotation symmetry:
\begin{equation}\begin{gathered}
\hat{U}_4 h^q(\vec{k}) \hat{U}^{-1}_4 = h^q(R_4\vec{k})\\
\hat{U}_4 = \text{diag}(e^{i3\pi/4},e^{i\pi/4},e^{-i\pi/4}, e^{-i3\pi/4}),
\label{eq:C4Def}\end{gathered}\end{equation}
where $R_4$ rotates $\vec{k}$ by $\pi/2$. In the presence of $C_4$ symmetry there are two topologically distinct phases: the QI, which occurs when $|\Delta_3|<2$, and a trivial insulator, which occurs when $|\Delta_3|>2$. 

To consider a continuum limit we note that a  transition between the two $C_4$ symmetric phases occurs at $|\Delta_3| = 2$, $\Delta_1 = \Delta_2 = 0$. When $\Delta_3 = 2$ a pair of 2D, two-component Dirac cones form at lattice momentum $\vec{k} = (\pi,\pi)$, and when $\Delta_3 = -2$ the Dirac cones form at $\vec{k} = (0,0)$. Without loss of generality let us restrict our attention to the band crossing where $\Delta_3 = 2.$ The low energy degrees of freedom near this critical point can be written in terms of Dirac fermions, with the continuum Lagrangian
\begin{equation}\begin{split}
\mathcal{L}_{\text{quad}} = \bar{\bm{\Psi}} [\gamma^0 i\partial_t + \gamma^1 i\partial_x + \gamma^2 i\partial_y + \bm{m}\cdot \bm{\tau}] \bm{\Psi}
\label{eq:QIContinuumLag}\end{split}\end{equation}
where $\bm{\Psi}$ and $\bar{\bm{\Psi}} = \bm{\Psi}^\dagger \gamma^0$ are 4 component spinors, and $\bm{m}\cdot \bm{\tau} = m_1 \tau^1 + m_2 \tau^2 + m_3 \tau^3$, with $m_1 \propto \Delta_1$, $m_2 \propto \frac{1}{\sqrt{2}}\Delta_2$, $m_3 \propto \frac{1}{\sqrt{2}}(\Delta_3-2)$. The $\gamma$ and $\tau$ matrices are defined as $\gamma^0 = \sigma^0 \otimes \sigma^3$, $\gamma^1 = i \sigma^3 \otimes \sigma^1$, $\gamma^2 = i \sigma^3 \otimes \sigma^2$, $\tau^1 = -\sigma^2\otimes \sigma^3$, $\tau^2 = -\sigma^1\otimes \sigma^3$, $\tau^3 = -\sigma^3\otimes \sigma^0$. In this form, it is manifest that the Lagrangian has two SU$(2)$ subgroups. First, there is the SU$(2)$ generated by the $\gamma$ matrices, which we will refer to as the spin of the fermions. Second, there is the SU$(2)$ generated by the $\tau$ matrices, which we will refer to as the isospin of the fermions. Eq. \ref{eq:QIContinuumLag} therefore describes a pair of two-component 2D Dirac fermions with isospin coupled to a mass vector $\bm{m}$. 

For a translationally invariant system (constant $\bm{m}$), the mass terms $m_1$ and $m_2$ break $C_4$ rotation symmetry (see Eq. \ref{eq:C4Def}). Hence, based on the lattice model, we can identify the Lagrangian where $\bm{m} = (0,0,m_3)$ as the QI for $m_3 < 0$, and the trivial insulator for $m_3 > 0$. Both of these phases are described by a pair of 2 component Dirac fermions with opposite mass terms. Intuitively we can think of this as a bilayer system where the layers, which are indexed by $\tau^3,$ have opposite (integer) Hall conductance. In addition to the ``light'' fermions in Eq. \ref{eq:QIContinuumLag}, this theory also includes a pair of two component  ``heavy'' regulator fermions, which will be left implicit for brevity. In the lattice model, the heavy fermions correspond to the massive excitations near crystal momentum $\vec{k} = (0,0)$ (since we have specified $\Delta_3\sim 2$ we avoid the parameter regime of the model where these fermions become massless). The Lagrangian for the heavy fermions, is identical to Eq. \ref{eq:QIContinuumLag} but with fixed mass vector $(0,0,M)$, where $M<0$ is large compared to the other relevant energy scales. We note that in principle there are also heavy fermion sectors at $\vec{k}=(0,\pi), (\pi, 0)$ which are related to each other by $C_4$ symmetry, but they will not affect our analysis, and we ignore them.

We are interested in the response properties of the Lagrangian Eq. \ref{eq:QIContinuumLag} in the presence of a background gauge field $A$ and a non-constant $\bm{m}$, about which much is already known \cite{jaroszewicz1984induced,abanov2000topological,chamon2008irrational,grover2008topological}. If we allow for broken translation symmetry the space of $C_4$-symmetric compatible mass vectors $\bm{m}(\vec{x}) \equiv (m_{1}(\vec{x}),m_{2}(\vec{x}),m_{3}(\vec{x}))$ is expanded such that they must satisfy
\begin{equation}\begin{split}
\bm{m}(\vec{x})  = (-m_{1}(R_4\vec{x}),-m_{2}(R_4\vec{x}),m_{3}(R_4\vec{x})),\label{eq:MassC4Rot}\end{split}\end{equation}
where $\vec{x}$ refers to space coordinates, and $x_\mu$ to space-time coordinates. 
Assuming that $\bm{m}$ varies slowly over length scales $\propto |\bm{m}|^{-1}$, the fermions can be integrated out at one loop order, which leads to the topological response term 
\begin{equation}\begin{split}
\mathcal{L}_{\text{top}} = \frac{\epsilon^{\mu\nu\rho}}{8\pi}  \bm{n}\cdot(\partial_\mu \bm{n}\times \partial_\nu \bm{n}) A_\rho,
\label{eq:ResponseSkyrm}\end{split}\end{equation}
where $\bm{n} = (n_1,n_2,n_3)$ is defined such that $\bm{m} \equiv m \bm{n}$ and $|\bm{n}|^2 = 1$. The value of $m$ does not affect the topological responses we are interested in, and will be taken to be a constant. Here, $j^\mu_{\text{sky}} = \frac{1}{4} \epsilon^{\mu\nu\rho} \bm{n}\cdot(\partial_\nu \bm{n}\times \partial_\rho \bm{n})$ is the skyrmion density of $\bm{n},$ and the response indicates that charge is bound to these skyrmions, $j^\mu = \frac{1}{2\pi} j^\mu_{\text{sky}}$. 

For our purposes, it will be useful to switch to a gauge field description of the skyrmions of $\bm{n}$\cite{han2017skyrmions} by defining a local SU$(2)$ transformation $\Omega$, which rotates the unit vector $\bm{n}$ to an arbitrary constant unit vector $\bm{N}$ at each point in space: 
\begin{equation}\begin{split}
  \Omega^{-1}(x_\mu) \bm{n}(x_\mu) \cdot \bm{\tau} \Omega(x_\mu)= \bm{N}\cdot \bm{\tau}.
 \label{eq:MassRotationDef}\end{split}\end{equation}
The choice of $\bm{N}$ is inconsequential, and can be changed via a global isospin rotation. The transformation in Eq. \ref{eq:MassRotationDef} modifies the Lagrangian by rotating the mass vector $\bm{m} = m\bm{n}\rightarrow m\bm{N}$, and generating a covariant derivative, $D_\mu = \partial_\mu -i A_\mu -i \bm{b}_\mu \cdot \bm{\tau}$, where $b^i_\mu = \frac{i}{4}\Tr[\tau^i \Omega^{-1}\partial_\mu \Omega]$. Despite the fact that there are $3$ gauge fields, $b^i_\mu$, one for each generator $\tau^i$ of SU$(2)$, there is actually only a U$(1)$ gauge symmetry. The gauge symmetry corresponds to the local U$(1) \subset $SU$(2)$ isospin rotations that leave the mass vector $m \bm{N}$ invariant. In terms of the gauge fields, $\bm{b}_\mu$ we can rewrite Eq. \ref{eq:ResponseSkyrm} as
\begin{equation}\begin{split}
&\mathcal{L}_{\text{top}} = \frac{1}{2\pi}\epsilon^{\mu\nu\rho} b^N_{\mu} \partial_{\nu} A_{\rho},\phantom{==} b^N_\mu \equiv \bm{N}\cdot \bm{b}_\mu
\label{eq:ResponseGauge}\end{split}\end{equation}
where $b^N$ is the U$(1)$ gauge field, which corresponds to the aforementioned U$(1) \subset$SU$(2)$ gauge symmetry, and should be regarded as a background field that encodes the skymrions of $\bm{n}$. Here, $j^\mu =-\frac{1}{2\pi} \epsilon^{\mu\nu\rho} \partial_\nu b_\rho^N$, from which we see that charge is bound to the vortices of $b^N$.

\begin{figure}
\centering
\includegraphics[width=0.25\textwidth]{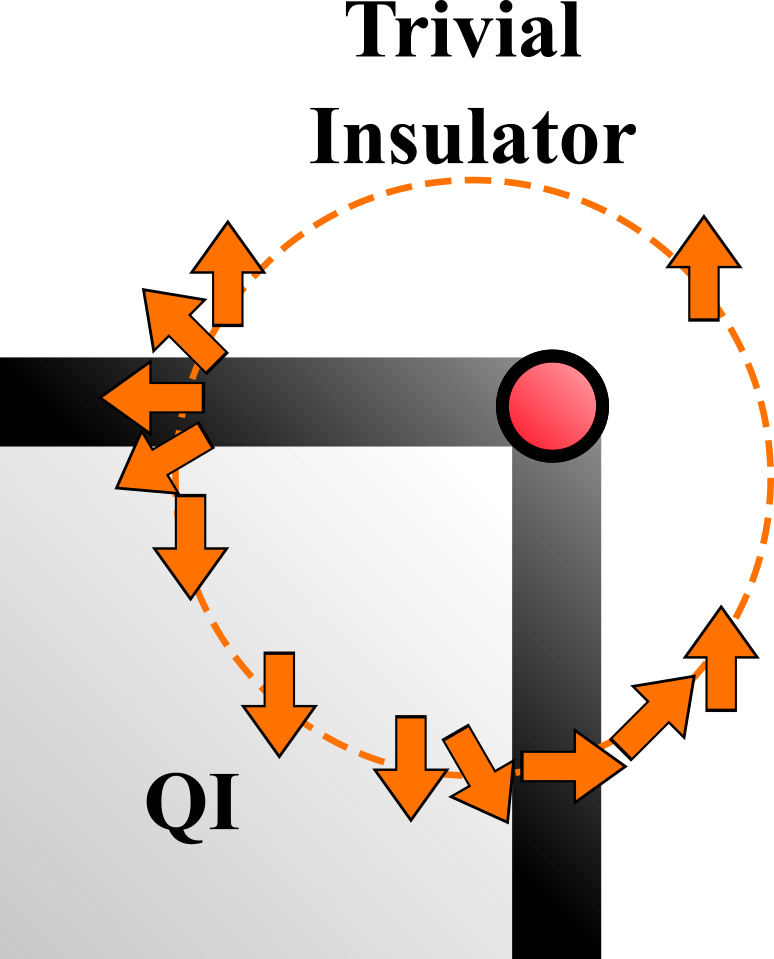}
\caption{A schematic of a $C_4$ symmetric domain corner between the QI and a trivial insulator. The arrows indicate the value of $\bm{n} = (0,\sin(\varphi),\cos(\varphi))$ on a loop that encircles the corner. At the domain walls, the value of $\hat{n}$ smoothly varies between the QI value of $\bm{n} = (0,0,-1)$, and the trivial value of $\bm{n} = (0,0,1)$.}\label{fig:DomainCornerDia}
\end{figure}

Now we are ready to consider the corner charge response for the $C_4$ symmetric QI. As we shall show, domain corners between the QI and trivial insulator correspond to $\pi$-vortices of $b^N$, and these $\pi$-vortices bind a half-integer charge. To see this, we consider a square sample of the QI embedded in a trivial insulator in a $C_4$ symmetric fashion. In the bulk of the QI $\bm{n} = (0,0,-1)$, while deep in the trivial region $\bm{n} = (0,0,1)$. The gapped domain walls between the QI and the trivial atomic insulator correspond to regions where $\bm{n}$ spatially interpolates between $(0,0,-1)$ and $(0,0,1)$. For simplicity, we will consider the situation where $n_1 = 0$ near the domain walls. This choice amounts to a choice of microscopic boundary conditions, and does not change the topological response. Since $n_1 = 0$ everywhere, it is convenient to switch to polar coordinates, $\bm{n} = (0,n_2,n_3) = (0,\sin(\varphi),\cos(\varphi))$, such that in the bulk of the QI $\varphi = \pi$, in the trivial region $\varphi = 0$, and at the domain walls $\varphi$ smoothly winds between $\pi$ and $0$. Based on Eq. \ref{eq:MassC4Rot}, $C_4$ symmetry requires that, $\varphi(\vec{x}) \rightarrow -\varphi(R_4\vec{x})$. So, $C_4$ symmetry requires that if domain walls normal to the $\pm x$-direction have $\varphi$ winding by $+\pi + 2q\pi $ ($q\in \mathbb{Z}$), then domain walls normal to the $\pm y$-direction must have $\varphi$ winding by $-\pi - 2q\pi $. Such a domain wall configuration is shown schematically in Fig. \ref{fig:DomainCornerDia}. If we rotate $\bm{n}$ according to Eq. \ref{eq:MassRotationDef}, $b^N_\mu = \frac{1}{2}\partial_\mu \varphi$, and the charge located at the domain corner is $Q_{\text{corner}} = -\frac{1}{2\pi}\oint b_i^N\cdot dl_i$ where the integral is over a loop that encircles the corner and is much larger than the width of the domain wall (see Eq. \ref{eq:ResponseGauge}). From our discussion above, $C_4$ symmetric domain walls generate a configuration where $\frac{1}{2\pi}\oint b_i^N\cdot dl_i = (q+\frac{1}{2})\pi$, and the corner charge is $Q_{\text{corner}} = 1/2 \mod(1)$\footnote{In this paper, we are only interested in the fractional part of the charge, as the total charge can always be shifted by an integer via the addition of local particles.}. We therefore find that the response theory correctly predicts the characteristic half-integer corner charge of the QI. In Appendix \ref{app:QIBound}, we also show the existence of these half-integer corner charges by considering the domain wall degrees of freedom of the continuum model.

\begin{figure}
\centering
\includegraphics[width=0.22\textwidth]{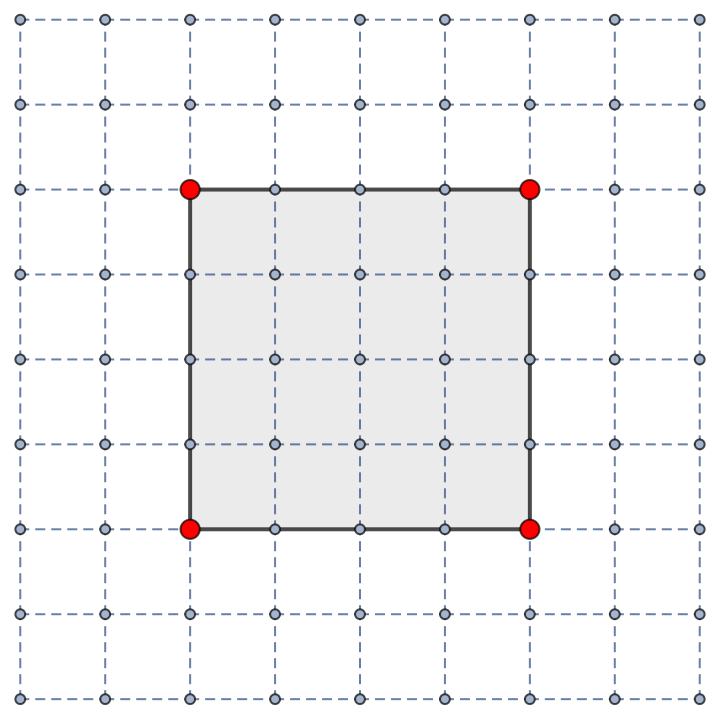}
\includegraphics[width=0.23\textwidth]{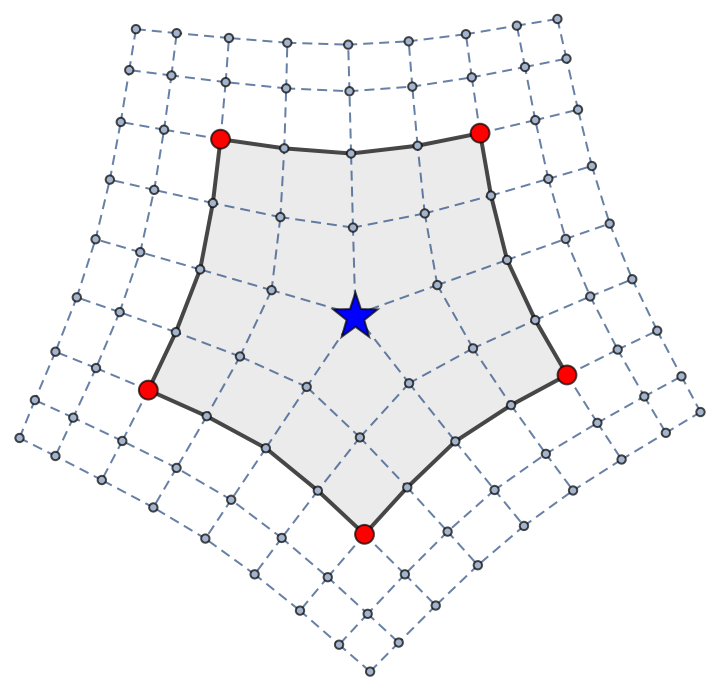}
\caption{Left: A sample (gray) without a disclination, which has 4 corners ($\color{red}\mathlarger{\bullet}$). Right: A sample with a $\pi/2$ disclination ($\color{blue}\mathlarger{\mathlarger{\mathlarger{\star}}}$), which has 5 corners.}\label{fig:LatticeDisc}
\end{figure}

Let us now move on to another notable feature of the QI: half-integer charge bound to $\pi/2$ disclinations of a $C_4$ symmetric lattice. It is useful to think of this effect as the electromagnetic response to singular sources of curvature, which are fluxes of $C_4$ symmetry. This response can be described by a Wen-Zee-like term\cite{wen1992shift,han2019generalized}. Before deriving the response term, it will be useful to first gain some intuition about the connection between the corner and disclination responses in rotation-invariant HOTIs\cite{li2020fractional}. To demonstrate this connection, we will consider a square sample of the QI embedded in a trivial insulator, as we previously discussed. The boundary of the QI traces out an angle of $2\pi$ with respect to the center of the sample, indicating that there are $4$ corners at which the orientation of the boundary changes by $\pi/2$. If a single disclination with Frank angle $\pi/2$ is added to the bulk of the QI, the boundary will instead trace out an angle of $5\pi/2$, and the sample must have $5$ corners (see Fig. \ref{fig:LatticeDisc}). Thus adding a disclination in the bulk requires the addition of an extra corner on the boundary. Furthermore, since the QI has half-integer corner charges, there will be an extra, anomalous, half-integer of charge at the boundary of the disclinated 5-corner sample. In order to have a total integer charge there must also be a half-integer charge bound to the disclination in the bulk. 
This argument indicates that charge conservation at the boundary is anomalous with respect to $C_4$ symmetry, since inserting a flux of $C_4$ symmetry into the bulk of the QI increases the charge at the boundary by a half-integer. This anomaly is canceled by the topologically non-trivial bulk of the QI. Here, we have only considered the $C_4$ symmetric QI, but it is straightforward to generalize this argument to other $C_n$ symmetric HOTIs.

The charge bound to disclinations can be interpreted as the bulk response of the QI, similar to how the charge bound to magnetic vortices is a bulk response of quantum Hall insulators. 
With this in mind, we now consider coupling the continuum theory in Eq. \ref{eq:QIContinuumLag} to curvature. Here we will consider a translationally invariant $C_4$ symmetric system ($\bm{m} = (0,0,m_3)$). The first observation we make is that in the continuum theory of the $C_4$ symmetric phases, the discrete $C_4$ rotation symmetry is enlarged to a continuous SO$(2)\simeq$U$(1)$ rotation symmetry. Under this symmetry, the spinors transform as $\bm{\Psi}\rightarrow \hat{U}(\theta) \bm{\Psi}$, where $\hat{U}(\theta) = \exp(i \theta [\frac{1}{2}\gamma^0 + \tau^3])$. The U$(1)$ rotations reduce to the $C_4$ rotations of Eq. \ref{eq:C4Def}, when $\theta = \pi/2$. It is worth noting that $\hat{U}(\theta)$ involves both the $\gamma$ matrices and the $\tau$ matrices, and therefore the spatial rotations rotate both the spin and the isospin of the fermions. 

To include curvature/disclinations in our description, we gauge this U$(1)$ rotation symmetry and introduce the background gauge field (spin connection) $\omega$\cite{lawrie2012unified}. The fermions in Eq. \ref{eq:QIContinuumLag} couple to curvature via a term proportional to the generator of spatial rotations in the covariant derivative: $D_\mu = \partial_\mu - i A_\mu - i\frac{1}{2}\omega_\mu \gamma^0 - i\omega_\mu\tau^3$. After integrating out the fermions, the response is given by the Wen-Zee term
\begin{equation}\begin{split}
\mathcal{L}_{\text{geo}} = \frac{\text{sgn}(m_3)-1}{2\pi}\epsilon^{\mu\nu\rho} \omega_\mu \partial_\nu A_\rho, 
\label{eq:GeoResponse}
\end{split}\end{equation}
where the addition of the $-1/2\pi$ term in the coefficient comes from the heavy fermions, which also couple to curvature. For the QI ($m_3 < 0$), $j^\mu = \frac{1}{\pi} \epsilon^{\mu\nu\rho} \partial_\nu \omega_\rho$, and $C_4$ disclinations of the lattice ($\pi/2$ vortices of $\omega$), bind a half-integer charge, while for the trivial phase ($m_3 > 0$) the Wen-Zee term vanishes. Eq. \ref{eq:GeoResponse} also indicates that the boundary of the QI is anomalous, with a mixed anomaly between spatial rotations and the U$(1)$ charge symmetry. The anomalous conservation law is
\begin{equation}\begin{split}
 \partial_\mu j^\mu |_{\text{boundary}} = -\frac{1}{\pi}\epsilon^{\mu\nu} \partial_\mu \omega_\nu|_{\text{boundary}},
 \label{eq:anaomalyBound}\end{split}\end{equation}
 where both sides of the equation are evaluated at the $1$D boundary of the QI. According to the integrated form of Eq. \ref{eq:anaomalyBound}, if we add $N_{\text{dis}}$ disclinations to the bulk of the QI, the charge localized at the boundary of the QI changes by $\Delta Q_{\text{boundary}} = -\frac{1}{2} N_{\text{dis}}$. Hence, we find that the amount of charge localized at the boundary is not conserved if $C_4$ symmetry is gauged, in agreement with our earlier argument.

Having separately discussed the continuum interpretations of the corner responses and the disclination responses, we are now in a position to present a unified description of both of these features. To do this, we first note that the spatial variation of a vector can be interpreted as the parallel transport of a constant vector  in the presence of an effective curvature. Indeed, if we consider a spatially varying mass vector $m(x_\mu)$, then using Eq. \ref{eq:MassRotationDef}, we find that
\begin{equation}\begin{split}
m_i(x_\mu+dx_\mu) = m_i(x_\mu)+ 2b^j_\nu \epsilon^{ijk} m_k(x_\mu) dx_\nu.  
\label{eq:MassParallel}\end{split}\end{equation}
Eq. \ref{eq:MassParallel} can be interpreted as the parallel transport of $\bm{m}$ with respect to an effective affine connection $2b^j_\mu \epsilon^{ijk}$ that encodes the effective curvature.  Importantly, this effective curvature couples to \textit{only} the isospin of the fermions, and leaves the spin of the fermions unaffected.  Let us now consider the effects of background curvature on $\bm{m}$. For a local patch of curved space, $\bm{m}$ satisfies
\begin{equation}\begin{split}
m_i(x_\mu+dx_\mu) = m_i(x_\mu)+ 2[b^j_\nu + \omega_\nu \delta^{j,3}] \epsilon^{ijk} m_k(x_\mu)dx_\nu,  
\label{eq:MassParallel2}\end{split}\end{equation}
where delta function, $\delta^{j,3}$, originates from the fact that the spin connection only couples to the isospin $\tau^3$. Therefore, the variation of $\bm{m}$ in curved space receives contributions from a combination of the physical background geometry $\omega$ and the effective geometry, $\bm{b}$.

We will now consider coupling the fermions in Eq. \ref{eq:QIContinuumLag} to background curvature, and allowing the mass vector $\bm{m} = m\bm{n}$ to vary. As before, we consider rotating the Lagrangian using $\Omega$, such that the theory has a constant mass vector $m \bm{N}$. After performing this rotation, the covariant derivative becomes $D_\mu = [\partial_\mu - i A_\mu  - i \omega_\mu \frac{1}{2} \gamma^0 - i \bm{b}_\mu \cdot \bm{\tau} - i \omega_\mu \bm{s}\cdot \bm{\tau}]$, where $s^j = \frac{1}{4}\Tr[\tau^{j}\Omega^{-1}\tau^3 \Omega]$ is the $j$-th projection of the isospin $\tau^3$ after the local rotation by $\Omega$. The theory therefore describes a system of fermions coupled to both physical curvature and the effective curvature (the latter induced by the variation of ${\bm{m}}$). After integrating out the fermions, the response theory is
\begin{equation}\begin{split}
&\mathcal{L}_{\text{full}} = \frac{\epsilon^{\mu\nu\rho}}{2\pi}\Big[b^N_\mu + \mathscr{s}\omega_\mu\Big]\partial_\nu A_{\rho},
\label{eq:ResponseFull}\end{split}\end{equation}
where $\mathscr{s} = (\bm{N}\cdot\bm{s}-1)$. The second term in $\mathscr{s}$ is from the heavy regulator fermions, which are coupled to the physical geometry, but, importantly, are \textit{not} rotated by $\Omega$ or coupled to the effective geometry. This response equation is a central result of this paper and indicates that charge is bound to fluxes of the combination $b^N+\mathscr{s}\omega$. As a consistency check, we note that when $\bm{m} = (0,0,m_3)$, $\mathscr{s} = \text{sgn}(m_3)-1$, in agreement with Eq. \ref{eq:GeoResponse}. It is worth noting that generically $\mathscr{s}$ is spatially dependent, and it is possible for $\mathscr{s} \omega$ to have a non-vanishing flux even when the flux of $\omega$ vanishes. Indeed, fluxes of $\mathscr{s} \omega$ can naturally appear on boundaries, far from any bulk disclinations, as we shall show below.

Previously, we pointed out that adding a disclination to the bulk of a QI with boundaries leads to an additional corner and its corresponding half integer corner charge. This behavior can be explicitly examined using Eq. \ref{eq:ResponseFull}. Let us start with a curvature free system ($\omega = 0$) where a domain wall located at $x=0$ separates a QI ($x<0$) and a trivial insulator ($x>0$)--we will implicitly assume that there are other domain walls located significantly far away such that the system has a global $C_4$ symmetry. As before, we shall consider boundary conditions such that $\bm{n} = (0,\sin(\varphi),\cos(\varphi))$ where $\varphi$ winds by $\pi + 2q\pi$ at the domain wall. In terms of the polar variables, $b^N_\mu = \frac{1}{2}\partial_\mu \varphi$, and $\mathscr{s} = \cos(\varphi)-1$. In the absence of curvature, there are no localized charges in this region of the domain wall, since $Q = \frac{1}{2\pi}\oint b^N_i dl_i = \oint \frac{1}{4\pi}\partial_i \varphi dl_i = 0$ for a loop far away from any other domain walls or corners. 

We will now add a $\pi/2$ disclination at $(x_0,y_0)$ which is deep in the bulk of the QI. To model this, we consider a local patch of space and choose $\omega$ as
\begin{equation}
\omega_x = 0, \phantom{=} \omega_y = \frac{\pi}{2}\Theta(x-x_0)\delta(y-y_0), \phantom{=} \omega_t =0,
\label{eq:discConfig}\end{equation}
where $\Theta$ is the step function. This configuration yields a point disclination around which there is a localized charge $Q_{\text{dis}}= \frac{1}{2\pi}\oint \mathscr{s} \omega_idl_i = 1/2$, where the integral is over a loop in the bulk of the QI that encircles $(x_0,y_0),$ and does not approach any domain walls. Let us now examine how this disclination affects the domain wall at $x=0$. From Eq. \ref{eq:discConfig} we see the curl of $\omega$ vanishes near the domain wall. However, $\mathscr{s}$ varies in space near the domain wall, and at $(x = 0,y_0)$ there is a charge $Q = \frac{1}{2\pi}\oint \mathscr{s} \omega_i dl_i = -1/2$ mod$(1)$, where the integral is over a loop that encircles $(x = 0,y_0)$ and does not approach the disclination or other domain walls. Based on our earlier discussion, we identify this charge as the charge bound to a new effective corner generated by the disclination\footnote{For a complete description of the corners in this curved geometry, one has to consider multiple local patches and framefields.}. We therefore see that Eq. \ref{eq:ResponseFull} correctly describes the charge bound to the disclination of the QI, as well as the additional charge localized at the boundary of the disclinated system.

We can also consider a ``fractional'' version of the response in Eq. \ref{eq:ResponseFull}, having a fractional prefactor $\nu$ that indicates a charge $\nu/2$ localized at corners or disclinations. Strictly speaking, such a fractional response is not properly quantized, and is more precisely expressed in terms of additional auxiliary gauge fields. We expect such a response to be realized in a \textit{fractional quadrupole insulator} (FQI), which, due to the fractional prefactor, must be a symmetry enriched topologically ordered phase\cite{lu2016classification}. 
Based on our earlier observation that the bulk of the QI can be treated as a bilayer system where the layers have opposite (integer) Hall conductances, we expect that the bulk of the FQI can be realized in a fractional Chern insulator (FCI) bilayer where the FCIs have opposite (fractional) Hall conductances. To analyze this system we can use flux attachment, and treat the FCI bilayer as a system of fermions interacting with dynamical gauge fields\cite{fradkin2013field,sohal2018chern}. Near the phase transition between the trivial and topological phases, the FCI bilayer is described by Eq. \ref{eq:QIContinuumLag} with covariant derivative, $D_\mu = \partial_\mu + \frac{1}{2}a^+_\mu(I + \tau^3) + \frac{1}{2}a^-_\mu(I - \tau^3)$ and additional Chern-Simons terms
\begin{equation}\begin{split}
\mathcal{L}_{\text{CS}} = \sum_{\eta = \pm }&\frac{\epsilon^{\mu\nu\rho}}{2\pi}\left[ a^{\eta}_\mu \partial_\nu c^{\eta}_\rho - \eta2k c^{\eta}_\mu \partial_\nu c^{\eta}_\rho - c^{\eta}_\mu \partial_\nu A_\rho \right],\\
\end{split}\end{equation}
where $a^{\pm}$ and $c^{\pm}$ are dynamic gauge fields\cite{lee2018emergent},  $\tau^3$ labels the different layers, and the gauge fields with $+$($-$) superscript live on the top (bottom) layer.
In this description the $m_3$ term is an intra-layer perturbation that controls the transition between the topological and trivial phases each layer, and the $m_1$ and $m_2$ terms are inter-layer couplings. If we assume that the fermions of the FCIs transform according to Eq. \ref{eq:C4Def} under $C_4$ rotations, then we can identify the Lagrangian with $\bm{m} =(0,0,m_3)$ as the FQI for $m_3<0$ and a trivial insulator for $m_3>0$. Following our earlier discussions, the gauge field response after integrating out the fermions is,\begin{equation}\begin{split}
\mathcal{L}_{\text{frac}} &= \sum_{\eta = \pm}\frac{\epsilon^{\mu\nu\rho}}{4\pi}\left[\eta a^{\eta}_\mu + 2b^N_\mu +2\mathscr{s}\omega_\mu \right]\partial_\nu  a^{\eta}_\rho  +  \mathcal{L}_{\text{CS}} \\
&\rightarrow \frac{\epsilon^{\mu\nu\rho}}{2\pi(2k+1)}\left[b^N_\mu + \mathscr{s} \omega_\mu \right]\partial_\nu A_\rho,
\end{split}\end{equation}
where we have included the spin connection $\omega$, and in the last line we have integrated out the dynamical gauge fields using their equations of motion. This is the response of a FQI with $\nu = 1/(2k+1)$, and describes a system with charge $1/2(2k+1)$ bound to corners and disclinations. In Appendix \ref{app:FQIBound}, we identify these corner charges by considering the domain wall degrees of freedom of the FCI bilayer. Other fractions $\nu$ can be found by considering multi-component versions of this theory. Based on this, we conclude that realizing the FQI state in FCI bilayers is plausible, however a microscopic model and a more detailed analysis of the flux attachment procedure is still needed.

As a final comment we note that QIs which have broken $C_4$ symmetry, but preserve the product of $C_4$ rotations and spinful time-reversal symmetry, $C_4\mathcal{T}^s$ (as one might find by dimensionally reducing the $3$D helical hinge insulator of Ref. \onlinecite{schindler2018higher} to $2$D) have corner charge responses described by our formalism, but the relevant bulk defects are not disclinations, but would instead be generated by gauging $C_4\mathcal{T}^s.$ Hence our theory is not immediately applicable to such systems. We provide a brief discussion of this problem in Appendix \ref{app:QSHComp} and leave the rest to future work.

\begin{acknowledgements}
{\textit{Note added}}. During the preparation of this manuscript an independent work appeared with some overlapping results in Ref. \onlinecite{huang2021effective}. These works were carried out independently.

We thank H. Goldman, R. Sohal, and O. Dubinkin for helpful discussions. JMM is supported by the National Science Foundation Graduate Research Fellowship Program under Grant No. DGE - 1746047. TLH thanks the US Office of Naval Research (ONR) Multidisciplinary University Research Initiative (MURI) grant
N00014-20- 1-2325 on Robust Photonic Materials with
High-Order Topological Protection for support. 

\end{acknowledgements}

\bibliography{Crystalline_Responses.bib}
\bibliographystyle{apsrev4-1}
\widetext
\begin{appendix}
\section{Equivalence of different quadrupole insulator lattice models}\label{app:DiffQuad}
In this paper we realize the quadrupole insulator (QI) by considering a four band tight binding model, with Bloch Hamiltonian,
\begin{equation}\begin{split}
h_{\text{quad}}(\vec{k}) = &\sin(k_x)[\sigma^3 \otimes \sigma^2]  + \sin(k_y)[-\sigma^3 \otimes \sigma^1] + \Delta_1 [-\sigma^1 \otimes \sigma^0]\\
&+\frac{1}{\sqrt{2}}[\Delta_2 + \cos(k_x)-\cos(k_y)][\sigma^2 \otimes \sigma^0]+\frac{1}{\sqrt{2}}[\Delta_3 + \cos(k_x)+\cos(k_y)][\sigma^3\otimes \sigma^3].
\label{Aeq:QILatticeModel}\end{split}\end{equation}
where $\sigma^{1,2,3}$ are the Pauli matrices, and $\sigma^0$ is the $2\times 2$ identity. We are primarily interested in the $C_4$ symmetric phases of this model, which occur when $\Delta_1 = \Delta_2 = 0$, and $\Delta_3 \neq \pm 2$, which we identified as the QI for $|\Delta_3| < 2$, and a trivial insulator for $|\Delta_3|>2$. This Hamiltonian should be compared to the one considered in the original formulation of the QI in Ref \cite{benalcazar2017quantized}, 
\begin{equation}\begin{split}
h_{\text{quad-2}}(\vec{k}) = &\lambda_x\sin(k_x)[-\sigma^2 \otimes \sigma^3]  + \lambda_y\sin(k_y)[-\sigma^2 \otimes \sigma^1] + \delta [\sigma^3 \otimes \sigma^0]\\
&+[\gamma_x + \lambda_x\cos(k_x)][\sigma^1 \otimes \sigma^0] +[\gamma_y + \lambda_y\cos(k_y)][-\sigma^2\otimes \sigma^2].
\label{Aeq:QILatticeModel2}\end{split}\end{equation}
This model has $C_4$ symmetry when $\delta = 0$, $\gamma_x = \gamma_y$, and $\lambda_x = \lambda_y$, and $\gamma_x/\lambda_x = \gamma_y/\lambda_y \neq \pm 1$. In Ref. \cite{benalcazar2017quantized} the $|\gamma_x/\lambda_x| = |\gamma_y/\lambda_y| < 1$ phase was identified as the QI, while $|\gamma_x/\lambda_x| = |\gamma_y/\lambda_y| > 1$ was identified as the trivial insulator. To show that these two lattice Hamiltonians are equivalent, we will use the unitary transformation 
\begin{equation}\begin{split}
\hat{U} = \left(
\begin{array}{cccc}
 \frac{1}{2} (-1)^{3/8} & -\frac{1}{2} (-1)^{5/8} & \frac{1}{2} (-1)^{3/8} & -\frac{1}{2}
   (-1)^{5/8} \\
 -\frac{1}{2} (-1)^{7/8} & \frac{1}{2} (-1)^{1/8} & -\frac{1}{2} (-1)^{7/8} &
   \frac{1}{2} (-1)^{1/8} \\
 \frac{1}{2} (-1)^{5/8} & \frac{1}{2} (-1)^{3/8} & -\frac{1}{2} (-1)^{5/8} & -\frac{1}{2}
   (-1)^{3/8} \\
 \frac{1}{2} (-1)^{1/8} & \frac{1}{2} (-1)^{7/8} & -\frac{1}{2} (-1)^{1/8} & -\frac{1}{2}
   (-1)^{7/8} \\
\end{array}
\right).
\label{Aeq:UnitaryBasisChange}\end{split}\end{equation}
Acting on Eq. \ref{Aeq:QILatticeModel} with this unitary leads to 
\begin{equation}\begin{split}
\hat{U}h_{\text{quad}}(\vec{k})\hat{U}^{-1} = &\sin(k_x)[-\sigma^2 \otimes \sigma^3]  + \sin(k_y)[-\sigma^2 \otimes \sigma^1] + \Delta_1 [\sigma^3 \otimes \sigma^0]\\
&+[\frac{\Delta_3 + \Delta_2}{2} + \cos(k_x)][\sigma^1 \otimes \sigma^0] +[\frac{\Delta_3 - \Delta_2}{2} + \cos(k_y)][-\sigma^2\otimes \sigma^2].
\label{Aeq:QILatticeModelTrans}\end{split}\end{equation}
This Hamiltonian is equivalent to the one in Eq. \ref{Aeq:QILatticeModel2} with $\gamma_x = (\Delta_3+\Delta_2)/2$, $\gamma_y = (\Delta_3-\Delta_2)/2$, $\delta = \Delta_1$, and $\lambda_x = \lambda_y = 1$.

\section{Boundary physics description of the corner modes of the quadrupole insulator}\label{app:QIBound}

The low energy physics of the quadrupole insulator (QI) is described by the continuum Lagrangian 
\begin{equation}\begin{split}
\mathcal{L}_{\text{quad}} = \bar{\bm{\Psi}} [\gamma^0 i\partial_t + \gamma^1 i\partial_x + \gamma^2 i\partial_y + m_3\tau^3] \bm{\Psi},
\label{Aeq:QIContinuumLag}\end{split}\end{equation}
where $\gamma^0 = \sigma^0 \otimes \sigma^3$, $\gamma^1 = i \sigma^3 \otimes \sigma^1$, $\gamma^2 = i \sigma^3 \otimes \sigma^2$, and $\tau^3 = -\sigma^3\otimes \sigma^0$. For $m_3<0$, the Lagrangian describes the QI and for $m_3>0$, the Lagrangian describes a trivial insulator. Both the QI and the trivial insulator are $C_4$ symmetric, where the $C_4$ rotations act on the spinors via $\bm{\Psi} \rightarrow \hat{U}_4 \bm{\Psi}$ where
\begin{equation}\begin{split}
\hat{U}_4 = \text{diag}(e^{i3\pi/4},e^{i\pi/4},e^{-i\pi/4}, e^{-i3\pi/4}).
\label{Aeq:C4SpinorDef}\end{split}\end{equation}

Let us consider a domain wall between the QI and a trivial insulator, where $m_3 = m\text{sgn}(x)$ ($m>0$ is a constant). Compared to the domain walls we considered in the main body of the text, this one is \textit{not} smooth at length scales $\propto m^{-1}$, and at this domain wall, there will be a pair of counter propagating modes. These modes correspond to the single particle wave functions
\begin{equation}\begin{split}
&u_{+x,R} = \tfrac{1}{\mathcal{N}} \begin{pmatrix} -1&1&0&0 \end{pmatrix} e^{- |m x|+ i k_y y}, \phantom{==}u_{+x,L} = \tfrac{1}{\mathcal{N}}  \begin{pmatrix} 0&0&-1&1 \end{pmatrix} e^{- |m x| + i k_y y},
\label{Aeq:DWFermionsX}\end{split}\end{equation}
where $\mathcal{N}$ is the normalization factor. Similarly, when $m_3 = m\text{sgn}(-x)$, the domain wall fermions correspond to the wave functions
\begin{equation}\begin{split}
&u_{-x,R} = \tfrac{1}{\mathcal{N}} \begin{pmatrix} 1&1&0&0 \end{pmatrix} e^{- |m x|+ i k_y y}, \phantom{==}u_{-x,L} = \tfrac{1}{\mathcal{N}}  \begin{pmatrix} 0&0&1&1 \end{pmatrix} e^{- |m x| + i k_y y},
\label{Aeq:DWFermionsX2}\end{split}\end{equation}
 when $m_3 = m\text{sgn}(y)$, the domain wall fermions correspond to the wave functions
\begin{equation}\begin{split}
&u_{+y,R}= \tfrac{1}{\mathcal{N}}  \begin{pmatrix} i&1&0&0 \end{pmatrix} e^{- |m y|+ i k_x x}, \phantom{==}u_{+y,L} = \tfrac{1}{\mathcal{N}}  \begin{pmatrix} 0&0&i&1 \end{pmatrix} e^{- |my| + i k_x x},
\label{Aeq:DWFermionsY}\end{split}\end{equation}
and when $m_3 = m\text{sgn}(-y)$, the domain wall fermions correspond to the wave functions
\begin{equation}\begin{split}
&u_{-y,R}= \tfrac{1}{\mathcal{N}}  \begin{pmatrix} -i&1&0&0 \end{pmatrix} e^{- |m y|+ i k_x x}, \phantom{==}u_{-y,L} = \tfrac{1}{\mathcal{N}}  \begin{pmatrix} 0&0&-i&1 \end{pmatrix} e^{- |my| + i k_x x},
\label{Aeq:DWFermionsY2}\end{split}\end{equation}

Let us now consider embedding a $L \times L$ sized square sample of QI ($m_3<0$) in a trivial insulator ($m_3>0$). We will take define the domain walls via $m_3 = m-2m\Theta(L/2+x)\Theta(-L/2-x)\Theta(L/2+y)\Theta(-L/2-y)$, where $\Theta$ is the step function. For this configuration, the sign of $m_3$ sharply at the domain walls, and there will be fermionic modes which traverse the boundary of the QI. The Lagrangian for the $1$D boundary modes is given by
\begin{equation}\begin{split}
\mathcal{L}_{\text{edge}} = &\psi^\dagger_{+x,R} i\partial_t \psi_{+x,R} -  v \psi^\dagger_{+x,R} i\partial_z \psi_{+x,R} - \psi^\dagger_{+x,L} i\partial_t \psi_{+x,L} -  v \psi^\dagger_{+x,L} i\partial_z \psi_{+x,L}\\
&+\psi^\dagger_{+y,R} i\partial_t \psi_{+y,R} -  v \psi^\dagger_{+y,R} i\partial_z \psi_{+y,R} - \psi^\dagger_{+y,L} i\partial_t \psi_{+y,L} -  v \psi^\dagger_{+y,L} i\partial_z \psi_{+y,L}\\
&+\psi^\dagger_{-x,R} i\partial_t \psi_{-x,R} -  v \psi^\dagger_{-x,R} i\partial_z \psi_{-x,R} - \psi^\dagger_{-x,L} i\partial_t \psi_{-x,L} -  v \psi^\dagger_{-x,L} i\partial_z \psi_{-x,L}\\
&+\psi^\dagger_{-y,R} i\partial_t \psi{-y,R} -  v \psi^\dagger_{-y,R} i\partial_z \psi_{-y,R} - \psi^\dagger_{-y,L} i\partial_t \psi_{-y,L} -  v \psi^\dagger_{-y,L} i\partial_z \psi_{-y,L},\\
\label{Aeq:DWLag}\end{split}\end{equation}
where $\psi_{\pm x,R/L}$ are the right and left moving fermions which are located on the boundary normal to the $\pm x$-direction. These fermions correspond to the domain wall modes $u_{\pm x,R/L}$ in Eq \ref{Aeq:DWFermionsX} and \ref{Aeq:DWFermionsX2}. Similarly, the $\psi_{\pm y,R/L}$ fermions are located on the boundary normal to the $\pm y$-direction, and correspond to domain wall modes $u_{\pm y,R/L}$ in Eq \ref{Aeq:DWFermionsY} and \ref{Aeq:DWFermionsY2}. The coordinate $z \in [-L/2,L/2]$ corresponds to the direction transverse to each boundary. Due to the corners, we enforce the following boundary conditions: 
\begin{equation}\begin{split}
    &\psi_{+x,R/L}(z = L/2) = \psi_{+y,R/L}(z = -L/2),\phantom{==} \psi_{+y,R/L}(z = L/2) = \psi_{-x,R/L}(z = -L/2),\\ & \psi_{-x,R/L}(z = L/2) = \psi_{-y,R/L}(z = -L/2), \phantom{==} \psi_{-y,R/L}(z = L/2) = \psi_{+x,R/L}(z = -L/2).
\end{split}\end{equation}

Let us now consider gapping out the $1$D fermions in Eq. \ref{Aeq:DWLag} while preserving $C_4$ symmetry. This can be done with mass terms of the form $ \psi^\dagger_{\pm x,R} \psi_{\pm x,L} + h.c.$ and $ \psi^\dagger_{\pm y,R} \psi_{\pm y,L} + h.c.$. Based on Eq. \ref{Aeq:C4SpinorDef}-\ref{Aeq:DWFermionsY2}, the $C_4$ symmetry transforms the edge fermions bilinears as 
\begin{equation}\begin{split}
   \psi^\dagger_{+x,R} \psi_{+x,R/L} \rightarrow -\psi^\dagger_{+y,R} \psi_{+y,R/L}\rightarrow \psi^\dagger_{-x,R} \psi_{-x,R/L} \rightarrow -\psi^\dagger_{-y,R} \psi_{-y,R/L} \rightarrow \psi^\dagger_{+x,R} \psi_{+x,R/L}.
\end{split}\end{equation}
 Therefore, the $C_4$ symmetric mass terms for the domain wall fermions take the form
\begin{equation}\begin{split}
\mathcal{L}_{\text{edge mass}} =  M\psi^\dagger_{+x,R} \psi_{+x,L} - M\psi^\dagger_{+ y,R} \psi_{+y,L} + M\psi^\dagger_{-x,R} \psi_{-x,L} - M\psi^\dagger_{-y,R} \psi_{-y,L} + h.c..
\label{Aeq:DWLagMass}\end{split}\end{equation}
Based on this, we see that at the corners the domain wall fermion mass changes sign. There will be Jackiw-Rebbi zero energy modes localized at the corners where the mass changes sign. These zero energy modes have half-integer charges, and corresponds to the characteristic half-integer corner charges of the QI.

\section{Boundary physics description of the corner modes of the fractional quadrupole insulator}\label{app:FQIBound}

The low energy physics of the fractional quadrupole insulator (FQI) is described by the continuum Lagrangian 
\begin{equation}\begin{split}
\mathcal{L}_{\text{frac-quad}} = &\bar{\bm{\Psi}} \Big[\gamma^0 [i\partial_t + \frac{1}{2}a^+_t(I + \tau^3) + \frac{1}{2}a^-_t(I - \tau^3)]  + \gamma^1 [i\partial_x + \frac{1}{2}a^+_x(I + \tau^3) + \frac{1}{2}a^-_x(I - \tau^3)]  \\ &\phantom{==}+ \gamma^2 [i\partial_y + \frac{1}{2}a^+_y(I + \tau^3) + \frac{1}{2}a^-_y(I - \tau^3)] + m_3\tau^3 \Big] \bm{\Psi}\\ & + \epsilon^{\mu\nu\rho}\left[\frac{1}{2\pi} a^{+}_\mu \partial_\nu c^{+}_\rho - \frac{2k}{2\pi} c^{+}_\mu \partial_\nu c^{+}_\rho - \frac{1}{2\pi} c^{+}_\mu \partial_\nu A_\rho + \frac{1}{2\pi} a^{-}_\mu \partial_\nu c^{-}_\rho +\frac{2k}{2\pi} c^{-}_\mu \partial_\nu c^{-}_\rho - \frac{1}{2\pi} c^{-}_\mu \partial_\nu A_\rho \right],
\label{Aeq:QIContinuumFracLag}\end{split}\end{equation}
where $\gamma^0 = \sigma^0 \otimes \sigma^3$, $\gamma^1 = i \sigma^3 \otimes \sigma^1$, $\gamma^2 = i \sigma^3 \otimes \sigma^2$, and $\tau^3 = -\sigma^3\otimes \sigma^0$, and $a^\pm$ and $c^{\pm}$ are \textit{dynamic} gauge fields ($A$ is a background gauge field). When $m_3<0$ the Lagrangian describes the FQI and when $m_3>0$, the Lagrangian describes a trivial insulator. This Lagrangian can be considered as a composite fermions description for a bilayer system, where each of the layers is near a phase transition between a trivial phase and a topological phase with the same topological order as a Laughlin state with filling $\pm 1/(2k+1)$. Both the FQI and the trivial insulator are $C_4$ symmetric, where the $C_4$ rotations act on the spinors via $\bm{\Psi} \rightarrow \hat{U}_4 \bm{\Psi}$ where
\begin{equation}\begin{split}
\hat{U}_4 = \text{diag}(e^{i3\pi/4},e^{i\pi/4},e^{-i\pi/4}, e^{-i3\pi/4}).
\label{Aeq:C4SpinorDefFrac}\end{split}\end{equation}

Let us consider a domain wall between the FQI and a trivial insulator, where $m_3 = m\text{sgn}(x)$ ($m>0$ is a constant). At this domain wall, there will be a pair of counter propagating modes. These modes correspond to the single particle wave functions, 
\begin{equation}\begin{split}
&u_{+x,R} = \tfrac{1}{\mathcal{N}} \begin{pmatrix} -1&1&0&0 \end{pmatrix} e^{- |m x|+ i k_y y}, \phantom{==}u_{+x,L} = \tfrac{1}{\mathcal{N}}  \begin{pmatrix} 0&0&-1&1 \end{pmatrix} e^{- |m x| + i k_y y},
\label{Aeq:DWFermionsXFrac}\end{split}\end{equation}
where $\mathcal{N}$ is the normalization factor. Similarly, when $m_3 = m\text{sgn}(-x)$, the domain wall fermions correspond to the wave functions
\begin{equation}\begin{split}
&u_{-x,R} = \tfrac{1}{\mathcal{N}} \begin{pmatrix} 1&1&0&0 \end{pmatrix} e^{- |m x|+ i k_y y}, \phantom{==}u_{-x,L} = \tfrac{1}{\mathcal{N}}  \begin{pmatrix} 0&0&1&1 \end{pmatrix} e^{- |m x| + i k_y y},
\label{Aeq:DWFermionsX2Frac}\end{split}\end{equation}
when $m_3 = m\text{sgn}(y)$, the domain wall fermions correspond to the wave functions,
\begin{equation}\begin{split}
&u_{+y,R}= \tfrac{1}{\mathcal{N}}  \begin{pmatrix} i&1&0&0 \end{pmatrix} e^{- |m y|+ i k_x x}, \phantom{==}u_{+y,L} = \tfrac{1}{\mathcal{N}}  \begin{pmatrix} 0&0&i&1 \end{pmatrix} e^{- |my| + i k_x x},
\label{Aeq:DWFermionsYFrac}\end{split}\end{equation}
and when $m_3 = m\text{sgn}(-y)$, the domain wall fermions correspond to the wave functions
\begin{equation}\begin{split}
&u_{-y,R}= \tfrac{1}{\mathcal{N}}  \begin{pmatrix} -i&1&0&0 \end{pmatrix} e^{- |m y|+ i k_x x}, \phantom{==}u_{-y,L} = \tfrac{1}{\mathcal{N}}  \begin{pmatrix} 0&0&-i&1 \end{pmatrix} e^{- |my| + i k_x x}.
\label{Aeq:DWFermionsY2Frac}\end{split}\end{equation}
Importantly, these domain wall modes still interact with the dynamic gauge fields, and are not particularly useful for describing the low energy physics. Instead, it is more useful to treat the FQI  as a bilayer system, where each layer has $\pm 1/(2k+1)$ topological order, and describe the edges in terms of a Luttinger liquid\cite{fradkin2013field}. The Lagrangian for such a Luttinger liquid is
\begin{equation}\begin{split}
\mathcal{L}_{\text{LL}} = \frac{2k+1}{4\pi} [\partial_y \phi_{R} \partial_t \phi_{R} - \partial_z \phi_{L} \partial_t \phi_{L}] - \frac{v}{4\pi}[(\partial_z \phi_{R})^2 + (\partial_z \phi_{L})^2] - \frac{g_f}{2\pi}\partial_z \phi_{R} \partial_z \phi_{L},
\label{Aeq:DWLLX}\end{split}\end{equation}
where $z$ is the direction tangent to the domain wall, and we have included a forwards scattering term $g_f$. The right and left moving local fermions of this theory correspond to the vertex operators $\exp(i (2k+1)\phi_{R})$ and, $\exp(-i (2k+1)\phi_{L})$ respectively. For appropriate large values of $g_f>0$ It is possible to gap out Luttinger liquid with a vertex interaction of the form $\cos((2k+1)[\phi_{R}+\phi_{L}])$. The charge density in the Luttinger liquid description is
\begin{equation}\begin{split}
\rho = \frac{1}{2\pi} \partial_y(\phi_{R}+\phi_{L}).
\label{Aeq:LLChargeX}\end{split}\end{equation}
When the Luttinger liquid acquires a gap from the vertex term, it is possible to have solitons where the expectation value of $\phi_{x,R}+\phi_{x,L}$ changes by $2\pi/(2k+1)$. These solitons correspond to the charge $1/(2k+1)$ anyons of the topologically ordered layers.

With this in mind, let us consider embedding an $L \times L$ sized square sample of FQI, where $m_3<0$, in a trivial insulator, where $m_3>0$. We will take define the domain walls via $m_3 = m-2m\Theta(L/2+x)\Theta(-L/2-x)\Theta(L/2+y)\Theta(-L/2-y)$, where $\Theta$ is the step function. At the boundary of the sample, there will be a $1$D Luttinger liquid. The Lagrangian for this Luttinger liquid can be written as 
\begin{equation}\begin{split}
\mathcal{L}_{\text{edge-LL}} = &\frac{2k+1}{4\pi} [\partial_z \phi_{+x,R} \partial_t \phi_{+x,R} - \partial_z \phi_{+x,L} \partial_t \phi_{+x,L}] - \frac{v}{4\pi}[(\partial_z \phi_{+x,R} )^2 + (\partial_z \phi_{+x,L})^2 ] - \frac{g_f}{2\pi}\partial_z \phi_{+x,R} \partial_z \phi_{x,L}\\
&+\frac{2k+1}{4\pi} [\partial_z \phi_{+y,R} \partial_t \phi_{+y,R} - \partial_z \phi_{+y,L} \partial_t \phi_{+y,L}] - \frac{v}{4\pi}[(\partial_z \phi_{+y,R})^2 + (\partial_z \phi_{+y,L})^2 ] - \frac{g_f}{2\pi}\partial_z \phi_{+y,R} \partial_z \phi_{+y,L}\\
&+\frac{2k+1}{4\pi} [\partial_z \phi_{-x,R} \partial_t \phi_{-x,R} - \partial_z \phi_{-x,L} \partial_t \phi_{-x,L}] - \frac{v}{4\pi}[(\partial_z \phi_{-x,R})^2 + (\partial_z \phi_{-x,L})^2] - \frac{g_f}{2\pi}\partial_z \phi_{-x,R} \partial_z \phi_{-x,L}\\
&+\frac{2k+1}{4\pi} [\partial_z \phi_{-y,R} \partial_t \phi_{-y,R} - \partial_z \phi_{-y,L} \partial_t \phi_{-y,L}] - \frac{v}{4\pi}[(\partial_z \phi_{-y,R})^2 + (\partial_z \phi_{-y,L})^2 ] - \frac{g_f}{2\pi}\partial_z \phi_{-y,R} \partial_z \phi_{-y,L}
\label{Aeq:DWLL}\end{split}\end{equation}
where $ \phi_{\pm x,R/L}$ are the right and left moving bosons at the boundaries normal to the $\pm x$ direction, $ \phi_{\pm y,R/L}$ are the right and left moving bosons at the boundaries normal to the $\pm y$ direction, and $z \in [-L/2,L/2]$ corresponds to the direction transverse to each boundary. Due to the corners, we enforce the following boundary conditions:  
\begin{equation}\begin{split}
  &\phi_{+x,R/L}(z = L/2) = \phi_{+y,R/L}(z = -L/2), \phantom{==}\phi_{+y,R/L}(z = L/2) = \phi_{-x,R/L}(z = -L/2),\\
  &\phi_{-x,R/L}(z = L/2) = \phi_{-y,R/L}(z = -L/2), \phantom{==} \phi_{-y,R/L}(z = L/2) = \phi_{+x,R/L}(z = -L/2).
\end{split}\end{equation}

Let us now consider gapping out the Luttinger liquid in Eq. \ref{Aeq:DWLL}, while preserving $C_4$ symmetry. As noted before, the Luttinger liquids can be gapped out via vertex terms of the form $\cos((2k+1)[\phi_{\pm x,R}+\phi_{\pm x,L}])$, and $\cos((2k+1)[\phi_{\pm y,R}+\phi_{\pm y,L}])$. In order to determine what vertex terms are allowed, we must determine how the bosonic fields transform under $C_4$ symmetry. Here, we shall make an assumption that the local vertex terms $\exp(i (2k+1)\phi_{\pm x,R})$ and $\exp(-i (2k+1)\phi_{\pm x,L})$ transform the same way under rotations as the domain wall modes $u_{\pm x,R}$ and $u_{\pm x,L}$ from Eq. \ref{Aeq:DWFermionsXFrac} and \ref{Aeq:DWFermionsX2Frac}, and that the vertex terms $\exp(i (2k+1)\phi_{\pm y,R})$ and $\exp(-i (2k+1)\phi_{\pm y,L})$ transform the same way under rotations as the domain wall modes $u_{\pm y,R}$ and $u_{\pm y,L}$ from Eq. \ref{Aeq:DWFermionsYFrac} and \ref{Aeq:DWFermionsY2Frac}. This means that for $C_4$ rotations, the bosonic fields transform as
\begin{equation}\begin{split}
&\phi_{+x,R} \rightarrow \phi_{+y,R} + \frac{\pi}{4(2k+1)}, \phantom{==} \phi_{+x,L} \rightarrow \phi_{+y,L} + \frac{3\pi}{4(2k+1)},\\
&\phi_{+y,R} \rightarrow \phi_{-x,R} + \frac{\pi}{4(2k+1)}, \phantom{==} \phi_{+y,L} \rightarrow \phi_{-x,L} + \frac{3\pi}{4(2k+1)},\\
&\phi_{-x,R} \rightarrow \phi_{-y,R} + \frac{\pi}{4(2k+1)}, \phantom{==} \phi_{-x,L} \rightarrow \phi_{-y,L} + \frac{3\pi}{4(2k+1)},\\
&\phi_{-y,R} \rightarrow \phi_{+x,R} + \frac{\pi}{4(2k+1)}, \phantom{==} \phi_{-y,L} \rightarrow \phi_{+x,L} + \frac{3\pi}{4(2k+1)}.
\end{split}\end{equation}
Based on this, the vertex operators which are compatible with $C_4$ rotations take the form 
\begin{equation}\begin{split}
\mathcal{L}_{\text{vertex}} = &\lambda \cos((2k+1)[\phi_{+x,R}+\phi_{+x,L}]) - \lambda \cos((2k+1)[\phi_{+ y,R}+\phi_{+y,L}])\\&  + \lambda \cos((2k+1)[\phi_{-x,R}+\phi_{- x,L}])  - \lambda \cos((2k+1)[\phi_{-y,R}+\phi_{-y,L}]).
\end{split}\end{equation}
To analyze the corner physics, let us restrict our attention to the boundaries normal to the $+x$ and $+y$ directions. If we take $\lambda<0$, then, in the strong coupling limit, $\phi_{+ x,R}+\phi_{+ x,L}$ acquires an expectation values equal to $2p\pi /(2k+1)$ for $p\in \mathbb{Z}$, and $\phi_{+y,R}+\phi_{+y,L}$ acquires an expectation values equal to $(2q+1)\pi /(2k+1)$ for $q\in \mathbb{Z}$. Due to the boundary condition $ \phi_{+x,R/L}(z = L/2) = \phi_{+y,R/L}(z = -L/2)$, at the corner there will be a soliton where the value of  $\phi_{+ x,R}+\phi_{+ x,L} \sim \phi_{+y,R}+\phi_{+y,L} $ changes by $(2q-2p+1)\pi /(2k+1)$. Based on Eq. \ref{Aeq:LLChargeX} this soliton has charge $(2q-2p+1) /2(2k+1)$. In particular, when $q = p$, the soliton charge is $1 /2(2k+1)$. This is the characteristic corner charge of the proposed FQI state. Corner charges with other values of $p$ and $q$ correspond to adding charge $1/(2k+1)$ anyons to the corners.

\section{Comparison between the $C_4$ symmetric quadrupole insulator, and $C_4\mathcal{T}^s$ symmetric quantum spin Hall insulator}\label{app:QSHComp}

In the main body of this paper, we focused on the quarupole insulator (QI) that is protected by $C_4$ symmetry. It is also worth commenting on a related HOTI with half-integer corner charges, which can be considered as a quantum spin Hall insulator (QSHI) where (spinfull) time reversal symmetry, $\mathcal{T}^{s}$, and $C_4$ symmetry are broken, but their product, $C_4\mathcal{T}^s$ is preserved. This system is essentially a dimensionally reduced version of the 3D chiral hinge insulator model from Ref. \cite{schindler2018higher}. The continuum Lagrangian for the QSHI is given by\cite{qi2008topological}
\begin{equation}\begin{split}
\mathcal{L}_{\text{QSH}} = \bar{\bm{\Psi}} [\gamma^0 i\partial_t + \gamma^1 i\partial_x + \gamma^2 i\partial_y + \bm{m}\cdot \bm{\tau}] \bm{\Psi}
\label{Aeq:QSHContinuumLag}\end{split}\end{equation}
where the $\gamma$ and $\tau$ matrices are defined as $\gamma^0 = \sigma^0 \otimes \sigma^3$, $\gamma^1 = i \sigma^3 \otimes \sigma^1$, $\gamma^2 = i \sigma^3 \otimes \sigma^2$, $\tau^1 = -\sigma^2\otimes \sigma^3$, $\tau^2 = -\sigma^1\otimes \sigma^3$, $\tau^3 = -\sigma^3\otimes \sigma^0$. This Lagrangian also describes the continuum limit of the QI. 

In the context of the QHSI, time-reversal is spinful, and given by $\mathcal{T}^{s} = \Gamma^2\Gamma^3\mathcal{K}$. Additionally, the QSHI has a $C_4$ rotation symmetry transforms the spinors via $\hat{U}^s_4 = \exp(i \frac{\pi}{4} \gamma^0)$ (for the QI, time reversal symmetry is given by $\mathcal{T}^{s} = \Gamma^2\Gamma^4\mathcal{K}$ and $C_4$ rotations transform the spinors via $\hat{U}_4 = \exp(i \frac{\pi}{4} \gamma^0 + \frac{\pi}{2}\tau^3)$. All the mass terms preserve the $C_4$ symmetry of the QSHI, while the $m_1$ and $m_2$ mass terms break $\mathcal{T}^s$. Based on this, we can identify the $\mathcal{T}^s$ symmetric Lagrangian with $\bm{m} = (0,0,m_3)$ as a QSHI for $m_3<0$ and a trivial insulator for $m_3>0$. We note that time reversal symmetry requires that $m_1=m_2=0$, even when translation symmetry is broken. Because of this, it is not possible to have $\mathcal{T}^s$ symmetric gapped domain walls between a QSHI and a trivial insulator. When $\mathcal{T}^{s}$ symmetry and $C_4$ symmetry are both relaxed, but their product, $C_4 \mathcal{T}^{s}$, is preserved, the $C_4 \mathcal{T}^{s}$ symmetric mass terms must satisfy
\begin{equation}\begin{split}
\bm{m}(\vec{x})  = (-m_{1}(R_4\vec{x}),-m_{2}(R_4\vec{x}),m_{3}(R_4\vec{x})).\label{Aeq:MassC4Rot}\end{split}\end{equation}
Based on Eq. \ref{Aeq:MassC4Rot} it is possible to have $C_4 \mathcal{T}^{s}$ symmetric gapped domain walls and corner charges by the exact same mechanism discussed in the main body of the text.

Physically, this system can be thought of as a QSHI that is coupled to a magnet with $C_4\mathcal{T}^s$ symmetry. A particularly simple way to do realize this is to couple the \textit{edges} of the QSHI to ferromagnets such that the ferromagnet have magnetization has the pattern $\bm{M}$ ($-\bm{M}$) on boundaries normal to the $\pm x$-directions ($\pm y$-directions). This configuration of magnetism breaks $C_4$ and $\mathcal{T}^s$ symmetry, but preserves $C_4\mathcal{T}^s$ symmetry, as desired. Due to the ferromagnets, the domain wall fermions of the QSHI are massive, and a half-integer of charge is bound to the corner, where the direction of the magnetization changes sign \cite{qi2008fractional,qi2008topological}. This analogy between the QI and the $C_4 \mathcal{T}^s$ QSHI provides an alternative physical platform to understand the corner physics we previously discussed. In this analogy, the $C_4$ symmetry fluxes ($C_4$ symmetry fluxes) of the QI correspond to $C_4 \mathcal{T}^{s}$ symmetry fluxes (combinations of disclinations and $\mathcal{T}^s$ symmetry fluxes) of the QSHI. Due to the addition of the $\mathcal{T}^{s}$ symmetry fluxes, this response cannot be described in terms of curvature, and one would have to instead consider gauging the magnetic point group symmetry $C_4 \mathcal{T}^{s}$. At this point, it is unclear if it is possible to describe the fluxes of $C_4 \mathcal{T}^{s}$ symmetry in terms of a $U(1)$ gauge field, the same way that it is possible to describe fluxes of $C_4$ symmetry in terms of the spin connection.

It also is useful to extend the analogy between the $C_4 \mathcal{T}^{s}$ QSHI and the QI to the fractional quadrupole insulator (FQI) we discussed in the main body of the text. Here, the analogous system to the FQI is a fractional quantum spin Hall insulator (FQSHI)\cite{bernevig2006quantum,levin2009fractional} where, as before, $\mathcal{T}^s$ and $C_4$ symmetry are broken, but their product, $C_4 \mathcal{T}^{s}$ is preserved.  A ($\mathcal{T}^s$ symmetric) FQSHI can be considered as a bilayer system composed of fractional quantum hall states at filling $\pm \nu$. For $\nu = 1/(2k+1)$, it is possible to use a flux attachment procedure to write the Lagrangian for the FQSHI as 
\begin{equation}\begin{split}
\mathcal{L}_{\text{FQSH}} = &\bar{\bm{\Psi}} \Big[\gamma^0 [i\partial_t + \frac{1}{2}a^+_t(I + \tau^3) + \frac{1}{2}a^-_t(I - \tau^3)]  + \gamma^1 [i\partial_x + \frac{1}{2}a^+_x(I + \tau^3) + \frac{1}{2}a^-_x(I - \tau^3)]  \\ &\phantom{==}+ \gamma^2 [i\partial_y + \frac{1}{2}a^+_y(I + \tau^3) + \frac{1}{2}a^-_y(I - \tau^3)] + m_3\tau^3 \Big] \bm{\Psi}\\ & + \frac{1}{2\pi} a^{+}d c^{+} - \frac{2k}{2\pi} c^{+}d c^{+} - \frac{1}{2\pi} c^{+}d A + \frac{1}{2\pi} a^{-}d c^{-} +\frac{2k}{2\pi} c^{-}d c^{-} - \frac{1}{2\pi} c^{-}d A,
\label{Aeq:QSHContinuumFracLag}\end{split}\end{equation}
where $a^\pm$ and $c^\pm$ are the dynamic gauge that implement flux attachment. This is the same Lagrangian we used to describe the FQI in the main text. In the context of the FQSHI, time reversal symmetry correspond to $\mathcal{T}^{s} = \Gamma^2\Gamma^3\mathcal{K}$, and there is a $C_4$ rotation symmetry transforms the spinors via $\hat{U}^s_4 = \exp(i \frac{\pi}{4} \gamma^0)$. Following the same logic as before, when $\mathcal{T}^{s}$ and $C_4$ symmetries are broken, but their product $C_4 \mathcal{T}^{s}$, the symmetry preserving mass terms $\bm{m}$ must satisfy Eq. \ref{Aeq:MassC4Rot}. Following the same logic as used in the main text, this leads to a system with charge $\nu/2$ localized at domain corners. 

Similar to the $C_4\mathcal{T}^s$ symmetric QSHI discussed above, $C_4\mathcal{T}^s$ symmetric FQSHI can be physically realized by coupling a $\mathcal{T}^s$ symmetric FQSHI to a magnet with $C_4\mathcal{T}^s$--e.g. by coupling the edges of the FQSHI to ferromagnets with magnetization $\bm{M}$ ($-\bm{M}$) on boundaries normal to the $\pm x$-direction ($\pm y$-direction). In a material context, it may be possible to realize this state in a material with significantly strong spin-orbit coupling and interactions that is coupled to a magnet where the magnetic order parameter has $d_{x^2-y^2}$ symmetry.

\end{appendix}

\end{document}